# Integrating social capital with urban infrastructure networks for more resilient cities

Ariel Favier[1*], Christine Hedde-von Westernhagen[2*], Meghan Krieg[3*], Bhaskar Kumawat[4*]

[*] University of California, Irvine; Ecology and Evolutionary Biology, Irvine, CA, USA
[2] Eindhoven University of Technology; Department of Industrial Engineering and Innovation Sciences, Eindhoven, The Netherlands
[3] Colorado School of Mines; Department of Mechanical Engineering, Golden, CO, USA
[4] University of Michigan LSA; Ecology and Evolutionary Biology, Ann-Arbor, MI, USA
* These authors contributed equally

**Introduction**

More than half of the world's population now lives in urban environments, which concentrate services and infrastructure to satisfy the material needs of a growing number of inhabitants. The interdependencies between physical infrastructure systems are required for cities to function efficiently, but simultaneously expose cities to new hazards. Failures that emerge from one infrastructure system and cascade through these interdependencies are becoming larger and more frequent due to climate change and growing urban environments. Because of the uneven distribution of resources and basic services, cascade failures often exacerbate pre-existing socioeconomic inequalities. Human communities rely on both social capital and infrastructure services to prepare for, manage, and recover from these challenging scenarios, but the overlap between social and physical infrastructure creates unpredictable feedback dynamics. While prior research has focused on either social capital or physical infrastructure in urban disaster management, an integrative view of these two perspectives is seldom explored. On one hand, a high level of physical infrastructure development and top-down disaster management can relieve community members from needing to rely on each other. Alternatively, reliance on community members (horizontal social capital) has shown to be effective in providing immediate and flexible responses.

In this paper, the feedback mechanisms between the physical and social layers of different urban designs are identified and analyzed to optimize relief response. Methodologically, we identify cities with high accessibility that have undergone disasters. From these cities, we measure their physical and social resilience indicators before and after disaster as a means to evaluate the impact of accessibility on disaster relief and preparedness. We will supplement this empirical analysis with a simulation that captures a cascade failure/disaster through a multilayer infrastructure and social network model. Overall, this research predicts that by focusing on infrastructure investment, maintenance, and distribution at a local level, high-level management can then shift its focus from local to regional resource distribution. This is because decentralization of infrastructure increases both accessibility and average proximity to key services, in turn improving governmental and community-based disaster response

The remainder of the paper is structured as follows. Section 1 and 2 review previous disaster-related work on physical infrastructure and social capital, respectively. Section 3 integrates social capital with physical infrastructure theoretically, and relates this to the concept of 15-minute cities. In Section 4, we lay out the methodology that will guide the next steps of this project, supplemented with graphical illustrations.

**Section 1: Physical Infrastructure**

Physical infrastructure like communication, transit, road, electric, food service, and water systems dictate the quality, connectivity, and flow of life in cities. While many physical infrastructure systems share the same space, these systems communicate with each other outside of their shared space through interdependencies. As an example, Buldyrev discussed how the single power station failure in Italy subsequently disrupted the internet network which spread to the 2003 nationwide blackout (Buldyrev, 2010). The power system relied on the internet network for control whereas the internet network depended on the power system for electricity for communication to subsequent power stations. Interdependencies within physical infrastructure characterize the extent to which a failure will propagate through all linked systems. These failures (such as the one described in Italy) are a product of natural phenomena, human error, random component failure,or system failure (Guo, 2017). The extent of the cascade/disaster determines the resilience of the critical infrastructure, its ability to "maintain critical function", and be adaptable in the event safeguards and response processes are improved (Wells, 2020).

Modeling cascade failures can predict infrastructure response to hazards which would allow us to uncover which safeguards or recovery methods must be further implemented to improve infrastructure resilience. Often, networks are used to model physical infrastructure because they both capture the structure and connectivity within a particular system and quantify interdependent connections between multiple systems (Wells, 2022). Although they are capable of simulating the complex feedback dynamics between multiple connected systems that lead to cascade failures, networks reduce physical infrastructure to a series of homogenous nodes and linear edges. Nodes may include subway stations where edges are the tracks or power stations where edges are power lines. The physical space where these infrastructure components (or nodes) reside is a crucial aspect to understand the robustness of these interdependent systems. In the event of a natural phenomenon, it is the location of the nodes not their connectivity that determines their vulnerability. Valdez discussed how a critical radius in space would describe how a failure in one node impacts any node within 'r' radius regardless of which infrastructure systems it belongs to (Valdez, 2020). In effect, this would recontextualize the network in space. But, because of the variable importance and spatial distribution of people and infrastructure nodes, a generalized critical radius does not truly capture how a cascade would behave in that system.

To explain, spatially a node may be identified by its physical coordinates and altitude, however the coupling of human design and natural environment uniquely changes the topography of each node in space. For example, a transportation hub in a hurricane prone region built in an underprivileged area with fewer access to critical resources behaves very differently during a cascade than a hub in an affluent area with minimal environmental risk. However, that same transportation hub in the affluent area may respond to failures in different infrastructure systems such as power that do not share the same spatial, componentry or environmental characteristics. In this sense, system resilience, that describes the compounding response and recovery between and within infrastructure systems, is quantified by how much functional loss systems experience from the same hazard (Wells, 2020). It is important to note that modeling cascade failures within infrastructure systems informs which resilience approaches are most appropriate (Wells, 2020). Thus, many interdependent infrastructure systems coupled with their natural environment and human design create hidden vulnerabilities that are not captured within a typical network model resulting in inaccurate evaluations of resilience.

Instead of modeling the entire overlapping network of physical infrastructure to understand resilience to hazards and disasters, Bruno refocuses on the individuals in urban environments impacted by events (Bruno, 2024). Bruno implies that access to critical resources dictate an individual's quality of life where infrastructure systems are just a means to deliver or provide access to those resources. If key resources such as hospitals, supermarkets are redistributed within a city such that most individuals are within 15 minutes of them, the impact of a disaster on individuals will shrink as less individuals will be deprived of critical needs. In this sense, the cascade failure through the physical infrastructure systems will be confined to the spatial, local level where the critical resources are concentrated. By improving accessibility to critical resources, we become more resilient to the complex interdependencies of physical infrastructure that give rise to systemic vulnerabilities.

While physical infrastructure is a crucial component to understanding an entire urban environment's resilience, it doesn't quantify the impact of social networks on both the structure and interconnectivity of a city and on the recovery, response and adaptability to disasters (cascade failures). In the 15 minute city model, Bruno discusses how the redistribution and restructuring of critical physical resources increases social equality, accessibility, and opportunities (Bruno, 2024). The strength of social relationships and behavior has the potential to create a more robust urban environment.

**Section 2: Social capital**

**Origin and theoretical considerations**

> "[Social capital] is the primary base on which a community response is built. In addition, social capital is the only form of capital, which is renewed and enhanced, quickly in emergency situations." (Dynes 2003, p.15)

Emergency situations that involve large-scale damage of critical infrastructure are stress tests not only for the physical components of a place but also for its social fabric. At the same time, as Dynes points out in the quote above, this social fabric appears to be the more durable one—often growing to its full potential under adverse conditions. Disaster research has long recognized this enormous potential of social capital, and the number of publications on the topic has increased especially through the integration with debates on climate change resilience (Meyer 2018).

The concept of social capital incorporates many aspects that determine the functioning of a society, not only in times of crisis. Nan Lin defines it as "resources embedded in a social structure which are accessed and/or mobilized in purposive actions." (Lin 1999, p.35). The concept dates back to social theorists like Bourdieu, Coleman, and Putnam, where it emerged in extension to the economic theory of capital as conceptualized by Karl Marx (Lin 1999, Portes 1998).

With its swift adoption across the social sciences, also the boundaries regarding the definition and operationalization of social capital have widened—not always to its benefit. While scholars agree that social capital is rooted in the relations and interactions of individuals, the most relevant theoretical divide concerns the functioning of social capital as an individual versus collective resource. The collective view, most prominently represented by Robert Putnam (Putnam 1993), has spurred criticism due to its tendency for tautological reasoning, which appears when identifying social capital as both the cause and effect of a phenomenon. For example: Cities with well functioning governments and a flourishing economy perform so well because of high social capital. At the same time, good government and economic development are used as indicators of high social capital (Portes 1998, p.19, referring to Putnam 1993).

While the possible circularity of the collective-resource perspective warrants caution, it is nonetheless empirically plausible to allocate social capital at both the individual and collective level if one sticks to social networks as being the foundation of the concept. For example, from the individual perspective, having high social capital can mean an abundance of social relationships which enhance closeness and accessibility to different resources (e.g., potential employers, skills and information, romantic partners). At the collective level, this can then manifest as high social capital if a large number of individuals can draw on such social relationships. However, the mere sum of individuals with high social capital is not sufficient for high collective social capital. The crucial determinant is the *distribution* of individual social capital in a population. High collective social capital then presents as a fairly even distribution of individual social capital across societal groups, such as socio-economic strata, or cultural and religious communities. From a network perspective, this would manifest as a so-called "small-world" network structure (Watts and Strogatz 1998), showing a high degree of clustering but also a substantial number of inter-cluster ties. This structure minimizes the average distance between individuals (maximizes accessibility) via inter-cluster ties, while maintaining a certain degree of social grouping that forms the basis of interpersonal trust. It is then also this collective social capital (expressed in distributional terms) that should be related to collective-level outcomes like economic performance, or disaster response.

**Social capital in disaster research**

When it comes to operationalizing social capital for empirical research, further dividing it into *bonding, bridging,* and *linking* social capital has emerged as the dominant typology (Meyer 2018). Bonding social capital is found in strong ties based mostly on kinship. Bridging capital goes beyond that and includes ties to other groups and communities, usually via co-membership in associations of place-based activities. Linking social capital, also termed vertical social capital, captures social ties to individuals who can facilitate access to resources which are not accessible via bonding or bridging ties (horizontal social capital), such as local or regional government officials.

In disaster situations, these different types–bonding, bridging, and linking social capital–also fulfill different functions (Aldrich and Meyer 2015, Panday et al. 2021). Bonding social capital is especially important at the onset of any disaster situation, as it provides for most people the first points of contact for material, informational, and emotional support (Hawkins and Maurer 2010, Shahid et al. 2022). In locally concentrated emergencies, kin- or friendship ties that are located in different areas may also provide shelter or access to other resources not available in the affected location (Garrison and Sasser 2009, Murphy 2007).

Bridging social capital is often deemed more crucial for disaster situations because it enables preparedness and management of challenges at the community level. Places with high bridging social capital have shown to exhibit better preparedness and faster recovery from disasters (Carpenter 2015, Ma et al. 2021). This is possible when a diverse set of necessary resources can be mobilized quickly and flexibly. This process is usually not centrally organized but emerges from the extension and recombination of community resources that often serve other purposes in non-emergency times (Murphy 2007). A prerequisite for the positive influence of bridging ties at a broader scale is a sufficient overlap and altruism between communities. Substantial disconnect between clusters of communities poses the danger of marginalizing certain communities due to distrust, antipathy, or simply lacking knowledge of their existence (Woolcock 1998). Bridging capital can play a crucial role especially in areas lacking disaster management from governments (Panday et al. 2021, Shahid et al. 2022).

Lastly, linking social capital connects local communities to governmental institutions. These can provide resources beyond what would be possible via community response, also in a more centralized manner. Furthermore, the extent of democratic participation, also in non-crisis times, impacts the effectiveness of government-led interventions by increasing knowledge and awareness of local needs and peculiarities at the government level (Pelling 1998, Reimer et al. 2013). This also highlights the role of linking social capital for the capacity of adaptation and organized development of disaster management in the long run (Hawkins and Maurer 2010). Again, the distribution of linking social capital should not be neglected. Communities that are systematically neglected in government investment will also have difficulty accessing government resources in disaster situations, potentially exacerbating their situation (Adger et al. 2001, Panday et al. 2021).

As will be discussed in the next section, it is the overall composition of a population's social network structure along these three different types of ties that we deem crucial for successful disaster preparedness and response. More importantly, we will show how social capital interacts with the previously discussed physical capital in the form of urban design and infrastructure, and suggest approaches to optimizing disaster preparedness and response accordingly.

**Section 3: Integrating the physical and the social**

**The overlapping roles of social and physical infrastructure in failure resilience**

As discussed previously, human communities rely on both social capital and infrastructural services to recover from challenging scenarios. These facets, and their effect on recovery have been explored extensively in previous literature but their interaction during the developmental stages of a community is largely ignored (Sanderson et al. 2022; Delilah Roque, Pijawka, and Wutich 2020; Aldrich 2012; Nakagawa and Shaw 2004; Danziger and Barabási 2022; Carpenter 2015). For example, people in a rural region with less access to infrastructure services like paved roads and/or reliable electricity are likely to rely more frequently on their social networks for services (Flora 1998). Alternatively, the development of social capital in a society with stable access to services like the internet may be hindered because of less incentive to socialize with local individuals (Young 2006; Neves 2013). Thus, it is important to address how sociality and the physical access to infrastructure grow together as communities emerge.

Infrastructure can directly affect how frequently and efficiently groups of people come together and shape the shared social capital. Public transport, for example, while allowing people to move between points of interest, can also bring people together by allowing them to meet other individuals more easily (Currie and Stanley 2008; Gray, Shaw, and Farrington 2006). Electricity and the internet work hand in hand to promote non-local interactions, allowing people in two different parts of a large city to be socially close together. However, these also decrease the likelihood of local interactions that are ultimately more important, especially in a disaster scenario (Young 2006; Neves 2013). The biggest role, however, is likely played by a pedestrian friendly infrastructure, promoting both a healthier lifestyle, and an easier exposure to third-spaces, i.e., places where people can meet and socialize outside of work and home (Littman 2022; Cabras and Mount 2017; Kim and Yang 2017).

Alternatively, a case can be made for social capital affecting infrastructural systems. The functioning of a local government, for example, is directly affected by social climate, and in turn affects the investment that a community is able to make in its infrastructure (Temkin and Rohe 1998; Riska Farisa, Prayitno, and Dinanti 2019). Resilience to disaster is in fact a direct result of this sort of investment both from the local and federal levels. On the other hand, narrowly-implemented yet strong social capital may also lead to a distrust of outsiders, which in turn can debilitate infrastructure due to overspecialization and a lack of diversity and skills in the population (Beyerlein and Hipp 2005). Frequent challenges like disasters or infrastructure failures are also likely to affect the feedback between social capital and physical infrastructure, which in turn drives recovery through these facets in the future. An understanding of the feedback between these components, and optimizing their interplay, is thus essential for resilience against disaster.

**Optimizing the feedback between physical and social infrastructure**

Optimizing for a positive feedback between social capital and infrastructure is necessary for the future well-being of a community. Even if the infrastructure in a place is lacking, local governments can again play a big role in promoting social cohesion through activities and community organization events (Wallis and Dollery 2002; Warner 2001; Rice 2001). At an even finer scale, a few households getting together to share and work on an activity can help build better social bonds, thus acting as a self-sustaining feedback loop. Regular maintenance and investment in infrastructure from a higher level of government can also help aid this interaction (Cooke 2010). In addition, subsidies and incentives to make a community more open to new talent can help bring in new diverse skills required for proper functioning of the infrastructure (Zhuang 2022; Bagnall et al. 2019).

A promising tool to analyze this interconnection between the infrastructure and social layers is to use the common language of networks (Fig. 1). Social ties are easily represented as a social network with edges denoting interaction between nodes that signify individuals. A large body of work has now been established in dealing with the structure and emergence of such social networks (Watts and Strogatz 1998; Newman 2006). Similarly, most physical infrastructure systems also take the form of a network. For example, the combination of electrical power plants and power lines is an example of a network that typifies the electrical grid. These physical networks connect to each other in different ways to create a multilayer structure comprising all different components of infrastructure. Because of the interdependencies inherent in such an assemblage, cascading failures are a frequent occurrence in such systems (Danziger and Barabási 2022; Brummitt et al. 2012; Schäfer et al. 2018; Krönke et al. 2020). However, the occurrence and frequency of such cascades between the social and physical layers of an extended physical-social network is unexplored.

Specifically, failures in the physical infrastructure can directly affect the rate at which individuals come in contact with each other, changing the connectedness of the social network. For example, interruptions to public transit and road networks decrease the ability of individuals to get together for social gatherings. Alternatively, a failure of the power grid may incentivize people to get out of their homes and interact. These effects can thus manifest as a change in the overall incidence of edges locally, as well as a reconfiguration of long-distance edges in a spatial social network. Flipping the direction of causality, a reduction in social interactions may bring down the connectedness of the social network and in turn affect how quickly infrastructure recovers in different areas of a city. A unidirectional recovery coupling of this sort has been studied for purely infrastructural networks in previous work (Danziger and Barabási 2022). However, analyzing the bi-directional feedback between social and physical systems is an important and novel facet that can add to our understanding of how communities become more resilient.

**15-minute cities and infrastructure resilience**

In light of the current ecological crisis, new ways of thinking of resilient and sustainable urban systems have been brought about (Da Silva et al. 2019). In particular, the framework of 15-minute cities (FMCs), first introduced in 2016 (Moreno et al. 2021; Moreno 2021), has taken hold in urban design, and several major cities around the world have begun efforts to implement it (Allam et al. 2022; Barbieri et al. 2023; Di Marino et al. 2022; Noworól et al. 2022). This framework integrates the concept of *chrono-urbanism*, the minimization of travel time to services as a means to improve quality of life, with the need for geographical proximity to generate accessible, inclusive, and sustainable cities (Rojas-Rueda et al. 2024). This approach strays from previous neighborhood-centered frameworks such as the Transit-Oriented Development (TOD) in the sense that it does not advocate for the redesign of neighborhoods around physically strict guidelines to increase proximity, but rather looks for the localization of services into pre-existing neighborhoods (Ibraeva et al. 2020). The FMC framework understands the neighborhood (and by extension, the city) as a complex, dynamic system, whose properties emerge and are not designed (Pozoukidou and Chatziyiannaki 2021), relying on community solidarity and participation as key components of neighborhood transformation (Moreno et al. 2021). Consequently, the ideal FMC is compact, and provides high access to essential services for a dense and heterogeneous population through mixed land use at the neighborhood level. In this sense, there could be a case for 15-minute cities as more optimized models of the feedback between the physical and social components of infrastructure, that rely on localized services for highly participative communities.

Simulating the rearrangement of services and amenities necessary for a diversity of cities around the world to become FMCs appears highly achievable in already densely populated areas, but nevertheless contingent on the preexisting layout (Bruno et al. 2024). Additionally, accounting for human mobility behavior, or the way humans move and interact with services and amenities over time, can help further optimize the way accessibility is increased (Zhang et al. 2022). Overall, the transformation into a FMC involves an increase in polycentricity, which in turn requires further development of transportation infrastructure (Lan et al. 2019). This has clear beneficial implications like the promotion of local employment (Duffy-Deno and Dalenberg 1993), traffic decongestion (Louf and Barthelemy 2014), and overall social capital (Östh et al. 2018), but can also increase long-term maintenance costs. Furthermore, the theoretical benefits of FMCs on social cohesion, mixing, and equality (Carpenter 2015) have been challenged (Casarin et al. 2023; Veneri and Burgalassi 2012). As more cities across the world adopt increased accessibility as a measure of resilience, it is necessary to study the effects of such an approach on disaster prevention, management, and recovery, and the relative contribution of social and physical infrastructure to an outcome. Consequently, we propose an analysis designed to directly test the effects of 15-minuteness on disaster resilience for future work.

**Section 4: Future work**

To evaluate the relationship between accessibility (i.e. 15-minuteness) and disaster resilience, we propose a global review of urban resilience indicators of cities exposed to major natural disasters in the last 20 years. First, we will collect all occurrences of major disasters using publicly available databases (EM-DAT: https://public.emdat.be/). Because there is a lack of consensus on the definition of a major disaster (Nations International Strategy for Di...), we will refer to major disasters as those affecting more than 100000 people, and we will restrict our search to floods and storms, which respectively account for 43.4% and 28.2% of total disasters (Barredo 2007). For a period of ten years centered on each event date, we will obtain city-level disaster resilience indicators belonging to six major categories (community planning and capacity building; economic recovery; health and social services recovery; housing recovery; infrastructure systems recovery; natural and cultural resources recovery, as described by Horney and collaborators (Horney et al. 2017) from publicly available databases (OECD: https://data-explorer.oecd.org/). We will adapt the methodology developed by Burton (Burton 2015) to build a composite indicator of disaster resilience at the city level, along with individual and category-based correlations of indicators, and we will correlate them with metrics of city accessibility (PT, F15; https://whatif.sonycsl.it/15mincity/). In particular, we will focus on the relative contribution of social and physical infrastructure resilience indicators in explaining the rate and extent of disaster recovery. We expect to see an overall positive correlation between accessibility and disaster resilience, expressed as a faster rate and larger extent of recovery of indicators after disasters to pre-disaster levels in more accessible cities. However, we expect cities that have adopted a FMC framework as part of their planning to display higher resilience than those of a similar accessibility that have not implemented it, due to the former having an improved social infrastructure, engagement and preparedness. Figure 1 summarizes the components of this analysis.

Building on these empirical insights, we will construct a conceptual multilayer network model that explicitly couples physical infrastructure and social capital of a stylized urban environment. We will then simulate disaster-induced disruptions of the physical infrastructure layer and analyze how this initial disruption affects physical accessibility of the city under varying social capital configurations and different infrastructure layouts. Figure 2 provides an overview of this analysis.

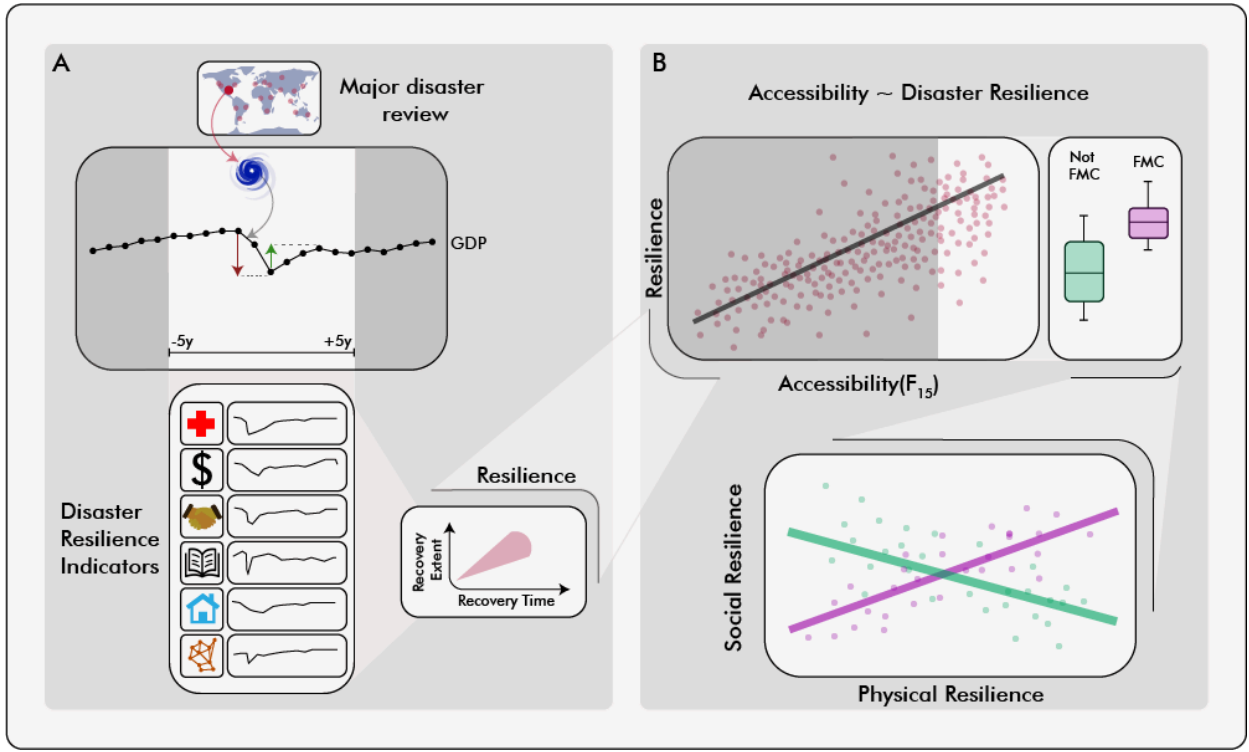

Figure 1: Evaluating the disaster resilience of 15-minute cities. (A) We will conduct a global review of major disasters and associated indicators of city resilience to obtain a condensed metric of disaster resilience based on the extent of recovery and the time taken for each city after a catastrophic event. (B) Afterwards, we will do a correlation analysis of disaster resilience and Accessibility (F15) of cities that had and had not adopted the FMC model guidelines (FMC/Not FMC) at the time of the catastrophic event, and will evaluate the relative contribution of indicators of social and physical resilience for cities with high accessibility in both categories.

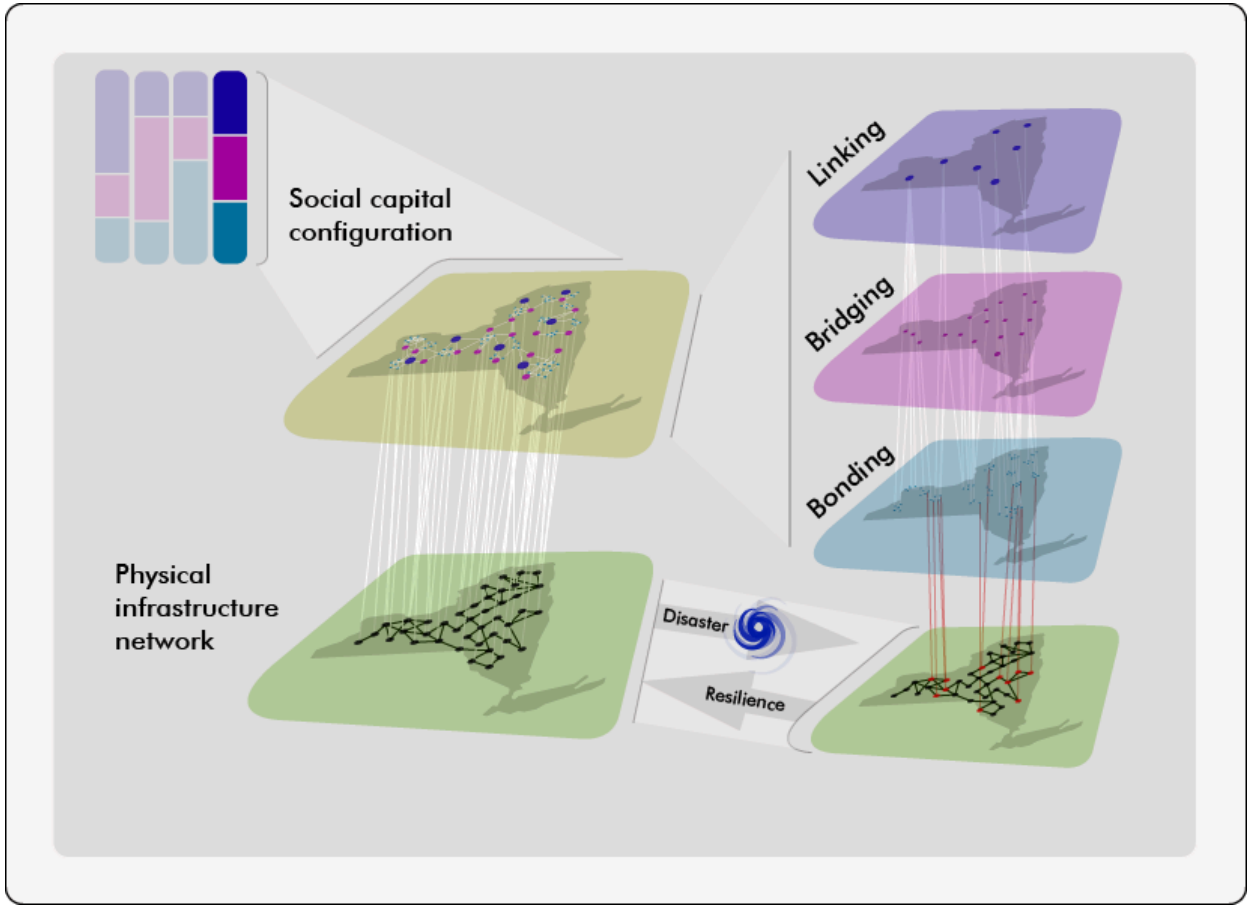

Figure 2: Simulating the resilience of different social capital configurations to disaster-induced disruptions of physical infrastructure. We show the social capital as a multilayered network of linking, bridging, and bonding ties. The social network is coupled with a layer of integrated physical infrastructure, where nodes represent points of interest (unspecified essential infrastructure and services). We propose to compare the time and extent of recovery after a simulated disruption in physical infrastructure for a city with different fractions of social ties, that compromises the initial accessibility of the affected areas (red nodes and links).